\documentclass[aps,prl,twocolumn,preprintnumbers, superscriptaddress,
               floatfix,nofootinbib]{revtex4-1}
\pdfoutput=1
\usepackage{graphicx}
\usepackage{epstopdf}
\usepackage{mathrsfs}
\usepackage{amssymb}
\usepackage{amsmath}
\usepackage{bm}
\usepackage{verbatim}
\usepackage{color}
\usepackage{multirow}

\def\stacksymbols #1#2#3#4{\def\theguybelow{#2}
    \def\vp{\lower#3pt}
    \def\sp{\baselineskip0pt\lineskip#4pt}
    \mathrel{\mathpalette\intermediary#1}}

\def\intermediary#1#2{\vp\vbox{\sp
     \everycr={}\tabskip0pt
     \halign{$\mathsurround0pt#1\hfil##\hfil$\crcr#2\crcr
              \theguybelow\crcr}}}

\def\be{\begin{equation}}
\def\ee{\end{equation}}
\def\bea{\begin{eqnarray}}
\def\eea{\end{eqnarray}}

\def\sp{\;\;\;,\;\;\;}

\def\lsim{\raise0.3ex\hbox{$\;<$\kern-0.75em\raise-1.1ex\hbox{$\sim\;$}}}
\def\gsim{\raise0.3ex\hbox{$\;>$\kern-0.75em\raise-1.1ex\hbox{$\sim\;$}}}

\def\inbar{\,\vrule height1.5ex width.4pt depth0pt}

\def\IC{\relax\hbox{$\inbar\kern-.3em{\rm C}$}}
\def\IQ{\relax\hbox{$\inbar\kern-.3em{\rm Q}$}}
\def\IR{\relax{\rm I\kern-.18em R}}
 \font\cmss=cmss10 \font\cmsss=cmss10 at 7pt
\def\IZ{\relax\ifmmode\mathchoice
 {\hbox{\cmss Z\kern-.4em Z}}{\hbox{\cmss Z\kern-.4em Z}}
 {\lower.9pt\hbox{\cmsss Z\kern-.4em Z}}
 {\lower1.2pt\hbox{\cmsss Z\kern-.4em Z}}\else{\cmss Z\kern-.4em Z}\fi}

\def\comment#1{}
\def\to{\rightarrow}

\def\u1x{U(1)_X}
\newcommand{\nc}{\newcommand}
\nc{\LL}{L}
\nc{\vv}{\tilde{v}}
\nc{\ccdot}{\!\cdot\!}
\nc{\gsm}{G_{SM}}
\nc{\vfive}{\mathbf{5}\oplus\mathbf{\overline{5}}}
\nc{\vten}{\mathbf{10}\oplus\mathbf{\overline{10}}}
\nc{\zhol}{Z^{\rm hol}}
\nc{\xfb}{\,{\rm fb}}

\begin{document}

\preprint{\tt KCL-PH-TH/2019-52}
\preprint{CERN-TH-2019-096}
\preprint{UT-19-14}
\preprint{ACT-04-19}
\preprint{MI-TH-1924}
\preprint{UMN-TH-3827/19}
\preprint{FTPI-MINN-19/18}

\vspace*{1mm}

\title{Cosmology with a Master Coupling in Flipped ${\rm SU}(5) \times {\rm U}(1)$: The $\lambda_6$ Universe}

\author{John Ellis}
\email{John.Ellis@cern.ch}
\affiliation{Theoretical Particle Physics and Cosmology Group, Department of
  Physics, King's~College~London, London WC2R 2LS, United Kingdom;\\
Theoretical Physics Department, CERN, CH-1211 Geneva 23,
  Switzerland;\\
National Institute of Chemical Physics and Biophysics, R\"{a}vala 10, 10143 Tallinn, Estonia}
\author{Marcos~A.~G.~Garcia}
\email{marcos.garcia@rice.edu}
\affiliation{Physics \& Astronomy Department, Rice University, Houston, TX
 77005, USA}

\author{Natsumi Nagata}
\email{natsumi@hep-th.phys.s.u-tokyo.ac.jp}
\affiliation{Department of Physics, University of Tokyo, 
Tokyo 113--0033,
Japan}

\author{Dimitri~V.~Nanopoulos}
\email{dimitri@physics.tamu.edu}
\affiliation{George P. and Cynthia W. Mitchell Institute for Fundamental
 Physics and Astronomy, Texas A\&M University, College Station, TX
 77843, USA:\\
Astroparticle Physics Group, Houston Advanced Research Center (HARC),
Mitchell Campus, Woodlands, TX 77381, USA:\\
Academy of Athens, Division of Natural Sciences, Athens, Greece}

\author{Keith A. Olive}
\email{olive@physics.umn.edu}
 \affiliation{
 William I.~Fine Theoretical Physics Institute, 
       School of Physics and Astronomy,
            University of Minnesota, Minneapolis, MN 55455, USA}

\begin{abstract} 

We propose a complete cosmological scenario based on a 
flipped ${\rm SU}(5) \times {\rm U}(1)$ GUT model that incorporates Starobinsky-like inflation,
taking the subsequent cosmological evolution carefully into account. 
A single master coupling, $\lambda_6$, connects the singlet, GUT Higgs and matter fields,
controlling 1) inflaton decays and reheating, 2) the gravitino
production rate and therefore the non-thermal abundance of the supersymmetric cold dark matter particle, 3) neutrino masses and 4) the baryon asymmetry of the Universe.

\end{abstract}

\maketitle




\setcounter{equation}{0}




It is common lore that the Universe may have been in a symmetric state soon after the Big Bang, but its subsequent evolution to the present-day universe with its content of matter, dark matter and neutrinos remains problematic. Typical grand unified theory (GUT) models require many seemingly unrelated couplings to explain
various physical observables. In this Letter we develop a complete cosmological scenario based on a detailed flipped SU(5)$\times$U(1) GUT
model~\cite{egnno2,egnno3} incorporating Starobinsky-like inflation \cite{staro}, and relate a host of cosmological observables through a single master coupling, denoted by $\lambda_6$. 

In addition to quark, lepton and Higgs fields, the model contains four gauge singlets that drive inflation,
provide a $\mu$-term for the mixing of the electroweak Higgs doublets, 
and a seesaw mechanism \cite{Minkowski:1977sc,Georgi:1979dq} for neutrino masses. Among the superpotential
couplings of the singlet fields there is one that couples the singlet, 
GUT Higgs fields and matter, denoted by $\lambda_6$. Remarkably this one coupling
controls 1) inflaton decays and therefore
the reheating temperature, 2) the gravitino
production rate and therefore the non-thermal abundance of the lightest supersymmetric particle (LSP) that is a candidate for cold dark matter, 3) neutrino masses, and 4) the baryon asymmetry of the Universe through leptogenesis \cite{fy}.
This Letter explores the deep correlations between these apparently disparate quantities that are all related by the master coupling $\lambda_6$---{\it the $\lambda_6$ Universe}.


In the flipped ${\rm SU}(5) \times {\rm U}(1)$ GUT
\cite{Barr,DKN,flipped2} motivated by string theory~\cite{AEHN}, all of the Standard Model (SM)
matter fields, as well as right-handed neutrinos, are embedded in
three generations of $\mathbf{10}~(1)$, $\bar{\mathbf{5}}~(-3)$, and $\mathbf{1}~(5)$
representations of SU(5), which are denoted by $F_i$, $\bar{f}_i$, and
$\ell^c_i$, respectively, where the numbers in the parentheses show the
U(1) charges in units of $1/\sqrt{40}$ and $i = 1,2,3$ are generation
indices. The representation assignments of the right-handed quarks and leptons are ``flipped'' with respect to those in
standard SU(5). The minimal supersymmetric standard model
(MSSM) Higgs fields $H_d$ and $H_u$ are in $\mathbf{5}~(-2)$ and
$\bar{\mathbf{5}}~(2)$ representations, denoted by $h$ and $\bar{h}$,
respectively. The GUT gauge group is broken into the SM gauge group by
$\mathbf{10}~(1)$ and $\overline{\mathbf{10}}~(-1)$ Higgs
representations of SU(5), which are denoted by $H$ and $\bar{H}$,
respectively. The four singlet chiral multiplets are denoted
$\phi_a$ ($a = 0, \dots, 3$), and we assume that the
inflaton can be identified with one of these, which we denote by $\phi_0$. 

The superpotential of this model~\cite{egnno2} is
given by 
\begin{align} \notag
W &=  \lambda_1^{ij} F_iF_jh + \lambda_2^{ij} F_i\bar{f}_j\bar{h} +
 \lambda_3^{ij}\bar{f}_i\ell^c_j h \\
&+ \lambda_4 HHh + \lambda_5
 \bar{H}\bar{H}\bar{h} + \lambda_6^{ia} F_i\bar{H}\phi_a 
\nonumber \\
&+ \lambda_7^a h\bar{h}\phi_a
 + \lambda_8^{abc}\phi_a\phi_b\phi_c + \mu^{ab}\phi_a\phi_b\,, 
\label{Wgen} 
\end{align}
where we impose a $\mathbb{Z}_2$ symmetry: $H\leftrightarrow
-H$, which forbids the mixing between the SM matter fields, Higgs color
triplets, and the Higgs decuplets, and suppresses the supersymmetric mass
term for $H$ and $\bar{H}$. Owing to the absence of these terms, 
rapid proton decay due to coloured Higgs exchange is avoided. In
addition, the doublet-triplet splitting problem is solved by the
missing-partner mechanism \cite{flipped2, Masiero:1982fe}. Without loss
of generality, we take $\lambda_2^{ij}$ and $\mu^{ab}$ to be
real and diagonal in what follows. Down-quark, up-quark and charged-lepton masses 
are related to the $\lambda_{1,2,3}$ couplings, respectively, and
(neglecting renormalization group effects for simplicity) 
$\lambda_2 \simeq \mathrm{diag} (m_u, m_c, m_t)/\langle \bar{h}^0\rangle$, where $\langle
\bar{h}^0\rangle$ is the weak scale vacuum expectation value (VEV) of $\bar{h}$.
A more detailed discussion of this model is given in
Ref.~\cite{egnno2}.

In order to describe cosmology, such a supersymmetric model must be embedded in a supergravity theory, 
which requires the specification of a K\"{a}hler potential $K$. 
In this model $K$ has the no-scale form~\cite{ekn2} that emerges from string
theory~\cite{Witten}. Denoting $\mu^{00} = m_s/2$ 
and assuming $\lambda_8^{000} = -
m_s/(3\sqrt{3} M_P)$, where $M_P \equiv (8\pi G_N)^{-1/2}$ is the
reduced Planck mass, the asymptotically-flat Starobinsky-like potential is
realized for the inflaton field $\phi_0$ \cite{ENO6}. The value $m_s
\simeq 3 \times 10^{13}$~GeV reproduces the measured value of the primordial power
spectrum amplitude~\cite{egnno2}. 

The inflaton $\phi_0$ couples directly to
the fields $F_i$ via the couplings $\lambda^{i0}_6$,
which play a central role in our analysis. Two other 
singlet fields, $\phi_1$ and $\phi_2$, also couple to $F_i$. 
The remaining singlet field does not couple to $F_i$, and develops a
supersymmetry breaking scale VEV which generates a $\mu$ mixing
term for the MSSM Higgs doublets. We assume that $\lambda_7^a = 0$
($a = 0,1,2$), so as to suppress $R$-parity violation.
This setup was introduced in Ref.~\cite{egnno2}, where it was called ``Scenario
B''. In this case, $R$-parity is violated in the
singlet sector, which is sufficiently sequestered from the observable
sector that the LSP has a lifetime
much longer than the age of the Universe \cite{egnno3}.


A general challenge in supersymmetric GUTs is the presence of multiple degenerate vacua \cite{supercosm,NOT}. While inflation might have left the Universe in the correct vacuum state, one should
follow the dynamic evolution of the universe, showing that the GUT phase transition occurred. Finite-temperature effects break the vacuum degeneracy through differences in the numbers of degrees of freedom
associated with the different phases~\cite{supercosm,NOT,Campbell:1987eb}. Although the global minimum generally lies in the symmetric state at temperatures of order the GUT scale, a GUT like SU(5) confines
at lower temperatures $T \sim 10^{10}$ GeV. This raises the GUT-symmetric vacuum energy, and opens the way towards successful cosmological evolution. 

GUT symmetry breaking in our model occurs along
one of the $F$- and $D$-flat directions in the scalar potential: a
linear combination of $\nu^c_H$ and $\nu^c_{\bar{H}}$, which are the SM
singlet components of $H$ and $\bar{H}$, respectively. We denote this
combination by $\Phi$, and call it the \textit{flaton}. Once $\Phi$
acquires a VEV, the ${\rm SU}(5) \times {\rm
U}(1)$ GUT gauge group is broken into the SM gauge group. The thirteen
Nambu-Goldstone chiral multiplets in $H$ and $\bar{H}$ are absorbed by
gauge multiplets, and the other six physical components are
combined with the triplet components in $h$ and $\bar{h}$ to make them
massive. The flat direction can be lifted by non-renormalizable
superpotential terms, e.g., of the form $W_{\text{NR}} \simeq
(H\bar{H})^4/M_P^{5}$. The flaton and flatino then obtain masses of order
the supersymmetry-breaking scale.

We focus on the portion of
parameter space where the {strong reheating} scenario discussed
in Ref.~\cite{egnno3} is realized. As shown in Ref.~\cite{egnno3}, in this case the
GUT symmetry is unbroken at the end of inflation. We further assume
that the system remains in the unbroken phase during reheating, 
as is confirmed in the following analysis. 
The phase transition is triggered by the difference in the number of
light degrees of freedom, $g$, between the broken and unbroken phases
\cite{supercosm,NOT,Campbell:1987eb, egnno2, egnno3}. Massless
superfields provide a thermal correction to the effective potential of $-
g\pi^2 T^4/90$, where $T$ denotes the temperature of the Universe. Since
the number of light degrees of freedom in the unbroken phase ($g = 103$)
is larger than that in the Higgs phase ($g = 62$), $\Phi$ is kept at the
origin at high temperatures. However, once the temperature drops below
 the confinement scale of the SU(5) gauge theory, $\Lambda_c$,
the number of light degrees of freedom significantly decreases ($g \leq
25$), and thus the Higgs phase becomes energetically favored
\cite{egnno2}. We have found that in this strong reheating scenario the incoherent component of the
flaton drives the phase transition if $\Lambda_c \gtrsim 2.3 (m_\Phi
M_{\rm GUT})^{1/2}$ \cite{egnno3}, where $m_\Phi$ and $M_{\rm GUT}$ are
the flaton mass and the GUT scale, respectively. For $m_\Phi = 10^4$~GeV and
$M_{\rm GUT} =10^{16}$~GeV, the above condition leads to $\Lambda_c
\gtrsim 2.3 \times 10^{10}$~GeV.

In the case of such strong reheating, 
the flaton decouples
from the thermal bath, and when $T \lesssim m_\Phi$ it becomes
non-relativistic and eventually dominates the energy density of the Universe until
it decays. The decay of the flaton generates a second period of
reheating. 
The amount of entropy
released by the flaton decay is estimated to be
\begin{equation}
 \Delta \simeq 1.6 \times 10^4\, \lambda^{-2}_{1,2,3,7} \,
\biggl(\frac{M_{\rm GUT}}{10^{16}~{\rm GeV}}\biggr)
\biggl(\frac{10~{\rm TeV}}{m^2_{\rm soft}/m_\Phi}\biggr)^{1/2} ~,
\end{equation} 
where $m_{\rm soft}$ stands for the typical value of sfermion
masses. 
It was shown in Ref.~\cite{egnno3} that if $|\lambda_{6}^{i0}| \gtrsim
{\cal O}(10^{-4})$, reheating is completed in the symmetric phase
via the dominant inflaton decay channel $\phi_0 \to F_i \bar{H}$. The reheating
temperature in this case is given by
\begin{equation}
 T_{\rm reh} \simeq 5.4 \times 10^{14} ~\mathrm{GeV}
\times \sqrt{10 \sum_{i} |\lambda_6^{i0}|^2} ~,
\label{eq:treh}
\end{equation}
indicating a direct relation between
$T_{\rm reh}$ and $\lambda_6$.

During reheating, gravitinos are produced via the
scattering/decay of particles in the thermal bath
\cite{weinberg,elinn,nos,ehnos,kl,ekn,eln,Juszkiewicz:gg,mmy,Kawasaki:1994af,
Moroi:1995fs,enor,Giudice:1999am,bbb,kmy,stef,Pradler:2006qh,ps2,rs,kkmy,
EGNOP}. For
the calculation of the gravitino production rate, we use the formalism 
outlined in \cite{rs}, but using the 
group theoretical factors and couplings
appropriate to flipped SU(5)$\times$U(1).

These gravitinos eventually decay into
LSPs, and the resultant ``non-thermal'' contribution to the LSP abundance
is given by
\begin{align}
 &\Omega_{\rm DM} h^2 \simeq  0.12  \,
\biggl(\frac{1.6 \times 10^4}{\Delta}\biggr)
\biggl(\frac{m_{\rm LSP}}{1~{\rm TeV}}\biggr)
\biggl(\frac{\sqrt{\sum_{i} |\lambda^{i0}_6|^2}}{0.0097}\biggr) \nonumber \\
 &=  0.12  \,
\biggl(\frac{1.6 \times 10^4}{\Delta}\biggr)
\biggl(\frac{m_{\rm LSP}}{1~{\rm TeV}}\biggr)
\biggl(\frac{T_{\rm reh}}{1.6 \times 10^{13} ~{\rm GeV}}\biggr)
~.
\label{eq:odm}
\end{align}
The total dark matter abundance is obtained by adding this non-thermal
component to the thermal relic density of the LSP,
which is reduced by a dilution factor
$\Delta$. Thus the LSP relic density
is also directly related to $\lambda_6$.


The neutrino mass structure in this model was studied in
Refs.~\cite{egnno2, egnno3}. 
As we noted above, only three singlet fields, including the inflaton, couple
to the neutrino sector.
The masses of the heavy states are approximately $(m_s, \mu_1, \mu_2)/2$,
and the mass matrix of the right-handed
neutrinos is obtained from a first seesaw mechanism:
\begin{equation}
 (m_{\nu^c})_{ij} = \sum_{a=0,1,2} \frac{\lambda_6^{ia} \lambda_6^{ja}}{\mu^a}
  \langle \tilde{\nu}_{\bar{H}}^c \rangle^2 ~,
\label{eq:mnuc}
\end{equation}
where we take $\langle \tilde{\nu}_{\bar{H}}^c \rangle = 10^{16}$~GeV in this
paper. We diagonalize the mass matrix (\ref{eq:mnuc})
using a unitary matrix $U_{\nu^c}$: $m_{\nu^c}^D = U_{\nu^c}^T m_{\nu^c}
U_{\nu^c}$. The light neutrino mass matrix is then obtained through a second
seesaw mechanism~\cite{Minkowski:1977sc,Georgi:1979dq}:
\begin{equation}
 (m_\nu)_{ij} = \sum_{k} \frac{\lambda_2^i \lambda_2^j (U_{\nu^c})_{ik}
  (U_{\nu^c})_{jk} \langle \bar{h}_0 \rangle^2 }{(m_{\nu^c}^D)_k} ~.
\label{eq:mnu}
\end{equation}
This mass matrix is diagonalized by a unitary matrix $U_\nu$ as $m_\nu^D
= U_\nu^* m_\nu  U_\nu^\dagger$. We note that, given a matrix
$\lambda^{ia}_6$, the mass eigenvalues of $m_\nu$ are uniquely
determined as functions of $\mu_1$ and $\mu_2$ 
via
Eqs.~\eqref{eq:mnuc} and \eqref{eq:mnu}.

On the other hand, as discussed in Ref.~\cite{Ellis:1993ks}, the PMNS
matrix differs from $U_\nu$ by an additional factor of a unitary matrix
$U_l^{}$: $U_{\rm PMNS} = U_l^{} U_\nu^\dagger$. This prevents us from
predicting the PMNS matrix in this framework. We note, however, that we
can instead use this equation to determine $U_\ell$ (given $U_{\rm
PMNS}$). It was found in~\cite{Ellis:1993ks} that the matrix $U_\ell$
affects the ratios between proton decay channels; for instance, $\Gamma
(p\to \mu^+ \pi^0)/\Gamma (p\to e^+ \pi^0) =
|(U_{l})_{12}|^2/|(U_l)_{11}|^2$, which is in general different from
the ratio predicted in an ordinary SU(5) GUT. A more detailed discussion of proton decay will be given in a forthcoming paper~\cite{egnno5}.

As can be seen from Eq.~\eqref{eq:mnuc}, right-handed neutrinos become
massive after $\bar{H}$ develops a VEV. In the strong reheating
scenario, therefore, right-handed neutrinos are massless and in thermal
equilibrium right after the reheating is completed. 
They
become massive and drop out of equilibrium almost instantaneously at the time
of the GUT phase transition
and eventually
decay non-thermally \cite{NOT, egnno3} to generate a lepton asymmetry
\cite{fy}. The lepton asymmetry is then converted to a baryon
asymmetry via the sphaleron process \cite{Kuzmin:1985mm}. The resultant
baryon number density is given by 
\begin{equation}
 \frac{n_B}{s} = -\frac{28}{79} \cdot \frac{135 \zeta (3)}{4\pi^4 g_{\rm
  reh}
  \Delta } \sum_{i=1,2,3} \epsilon_i ~,
  \label{nbs}
\end{equation}
where \cite{Ellis:1993ks, egnno3}
\begin{equation}
 \epsilon_i 
= 
\frac{1}{2\pi}\frac{ \sum_{j\neq i} {\rm Im} 
\left[\left(
U_{\nu^c}^\dagger (\lambda_2^D)^2 U_{\nu^c}
\right)_{ji}^2\right] }{\left[
U_{\nu^c}^\dagger (\lambda_2^D)^2 U_{\nu^c}
\right]_{ii}}
g\biggl(\frac{m^2_{\nu^c_{j}}}{m^2_{\nu^c_{i}}}\biggr) ~,
\end{equation}
with \cite{Covi:1996wh}
\begin{equation}
 g(x) \equiv -\sqrt{x} \biggl[
\frac{2}{x-1} + \ln \biggl(\frac{1+x}{x}\biggr)
\biggr] ~.
\label{eq:gx}
\end{equation}
 It is important to note that the sign in (\ref{nbs}) is 
 fixed: in order to obtain $n_B/s > 0$, we must require $\sum_{i} \epsilon_i < 0$.


As we see in Eqs.~\eqref{eq:treh} and \eqref{eq:odm}, the coupling
$\lambda_6$ determines the reheating temperature and the non-thermal
component of the dark matter abundance. This coupling also controls the
neutrino mass and baryon asymmetry through the right-handed neutrino
mass matrix in Eq.~\eqref{eq:mnuc}. 

We now investigate numerically the effect of the
$\lambda_6$ coupling on these physical observables.
To this end, we perform a parameter scan of $\lambda_6$. 
We first write it in the form $\lambda_6 = r_6 M_6$, where $r_6$ is a
real constant and $M_6$ is a complex $3\times 3$ matrix. We then scan
$r_6$ logarithmically over the range $(10^{-4},
1)$ choosing a total of 2000 values. For each value of $r_6$, we generate 2000 random complex $3\times 3$
matrices $M_6$ with each component taking a value of ${\cal O}(1)$. 

For each value of $\lambda_6$, we obtain the mass eigenvalues
of light neutrinos as functions of $\mu_1$ and $\mu_2$ as described above. We then
determine these two $\mu$ parameters by requiring that the observed
values of the squared mass differences are reproduced; namely, for the
normal ordering (NO) case, $m_2^2 - m_1^2 \equiv \Delta m_{21}^2 = 
7.39 \times 10^{-5}~{\rm eV}^2$ and $m_3^2 - m_1^2 \equiv \Delta
m_{31}^2 = 2.525 \times 10^{-3}~{\rm eV}^2$, and for the inverted
ordering (IO) case, $m_2^2 - m_1^2 = 7.39 \times 10^{-5}~{\rm eV}^2$ and
$m_3^2 - m_2^2 \equiv \Delta m_{32}^2 = - 2.512 \times 10^{-3}~{\rm
eV}^2$ \cite{nufit}.

We generate the same number of $\lambda_6$ matrices for each mass
ordering, and find solutions for 9839 and 730 matrix choices for
the NO and IO cases, respectively,
out of a total of $4 \times 10^6$ models sampled. This difference indicates that the NO case is favored in our
model. We find that the lightest neutrino mass is $\lesssim
10^{-5}$~eV in both cases. In the case of NO, the heavier
neutrinos have masses $\simeq \sqrt{\Delta m_{21}^2} = 8.6 \times
10^{-3}$~eV and $\simeq \sqrt{\Delta m_{31}^2} = 5.0 \times 10^{-2}$~eV. In
the IO case, on the other hand, both of the heavier states have masses
$\simeq \sqrt{|\Delta m_{32}^2|} = 5.0 \times 10^{-2}$~eV. The sum of
the neutrino masses is then given by $\sum_{i} m_{\nu_i} \simeq 0.06$~eV
and 0.1~eV for NO and IO, respectively. These predicted values are below
the current limit imposed by Planck 2018 \cite{Aghanim:2018eyx}, $\sum_i
m_{\nu_i} < 0.12$~eV, but can be probed in future CMB experiments such
as CMB-S4 \cite{Abazajian:2016yjj}. Moreover, the IO case can be probed
in future neutrino-less double beta decay experiments, whereas testing the
NO case in these experiments is quite challenging \cite{Agostini:2017jim}.

\begin{figure}
{\includegraphics[width=0.43\textwidth]{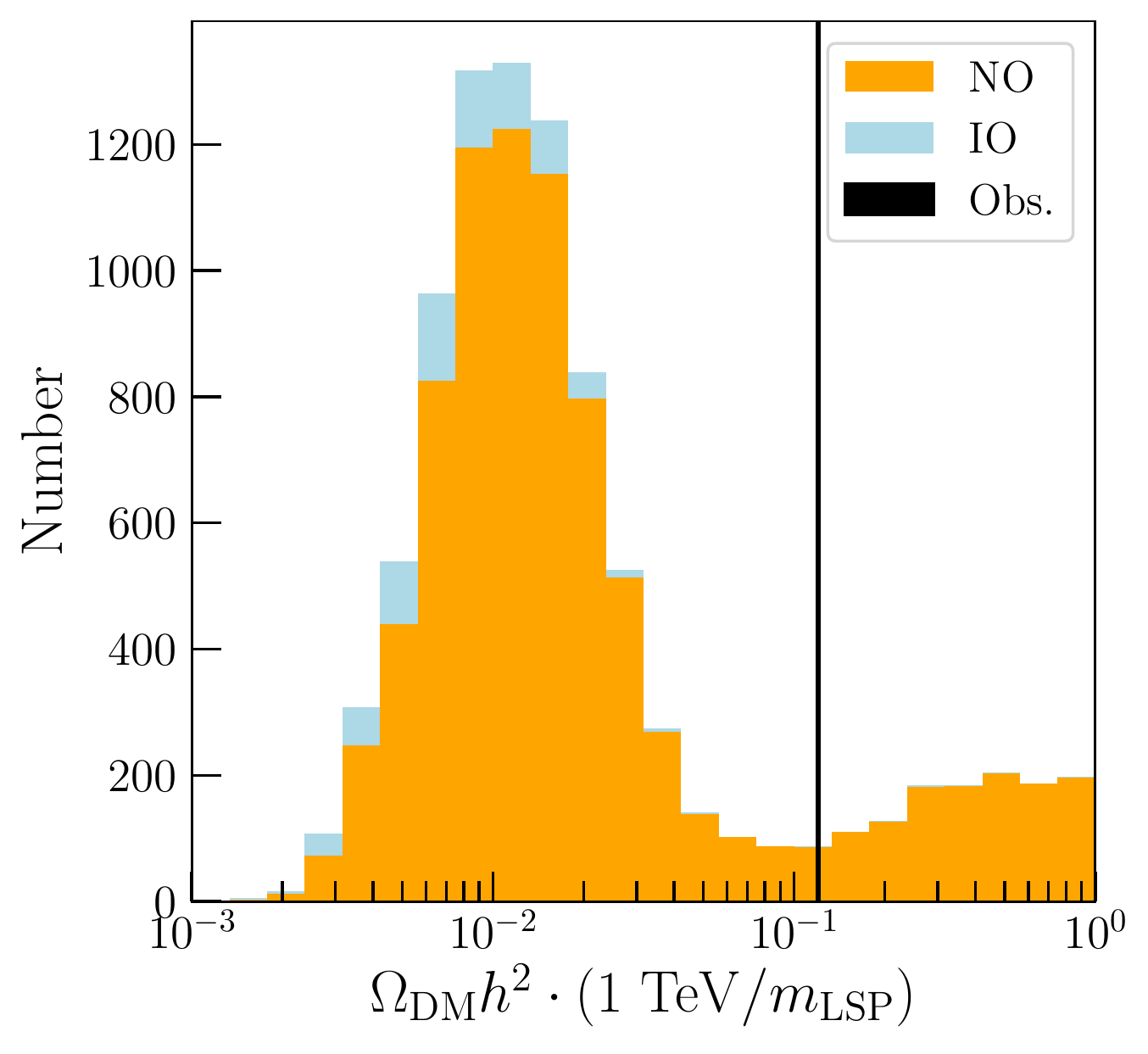}} 
\caption{
Histogram of values of $\Omega_{\rm DM} h^2$ that results from the parameter scan of $\lambda_6$,
assuming $m_{\rm LSP} = 1$~TeV, for the NO and IO cases.
}
\label{fig:omdm}
\end{figure}

We show in Fig.~\ref{fig:omdm} the distribution of the non-thermal dark matter density produced by gravitino decays in these solutions for $\lambda_6$. We find that many parameter solutions predict 
$\Omega_{\rm DM} h^2 \simeq 10^{-2}$ for $m_{\rm LSP} = 1$~TeV, corresponding to $T_{\rm
reh} \simeq 10^{12}$~GeV (see Eq.~\eqref{eq:odm}), while some solutions yield $\Omega_{\rm DM} h^2 \simeq 10^{-1}$ corresponding to
a reheating
temperature as high as $T_{\rm reh} \simeq 10^{13}$~GeV. In both cases,
the reheating temperature is much higher than the SU(5) confinement
scale $\Lambda_c$, satisfying the strong reheating condition
\cite{egnno3}.

\begin{figure}
{\includegraphics[width=0.43\textwidth]{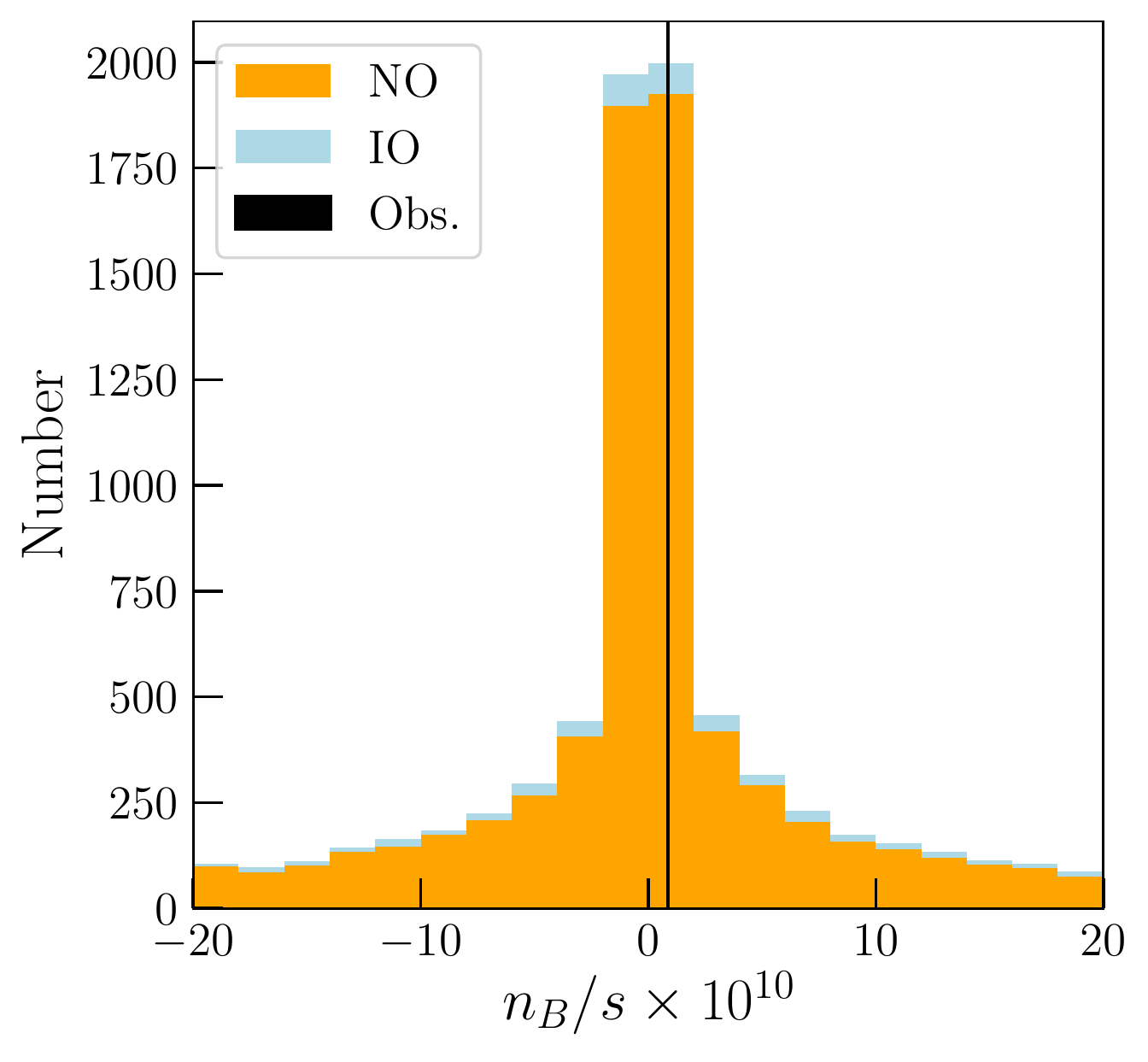}} 
\caption{
Histogram of values of $n_B/s$ for $\Delta = 10^4$. The vertical black
 solid line shows the observed value.
}
\label{fig:bau}
\end{figure}

In Fig.~\ref{fig:bau} we show the distribution of $n_B/s$ for $\Delta =
10^4$, where we see that both positive and negative baryon asymmetries can
be obtained. In particular, the observed value (in both magnitude and sign) of the baryon asymmetry
$n_B/s = 0.87 \times 10^{-10}$ \cite{Aghanim:2018eyx}, which is shown as
the vertical solid line, can easily be explained in our scenario. 

\begin{figure}
{\includegraphics[width=0.43\textwidth]{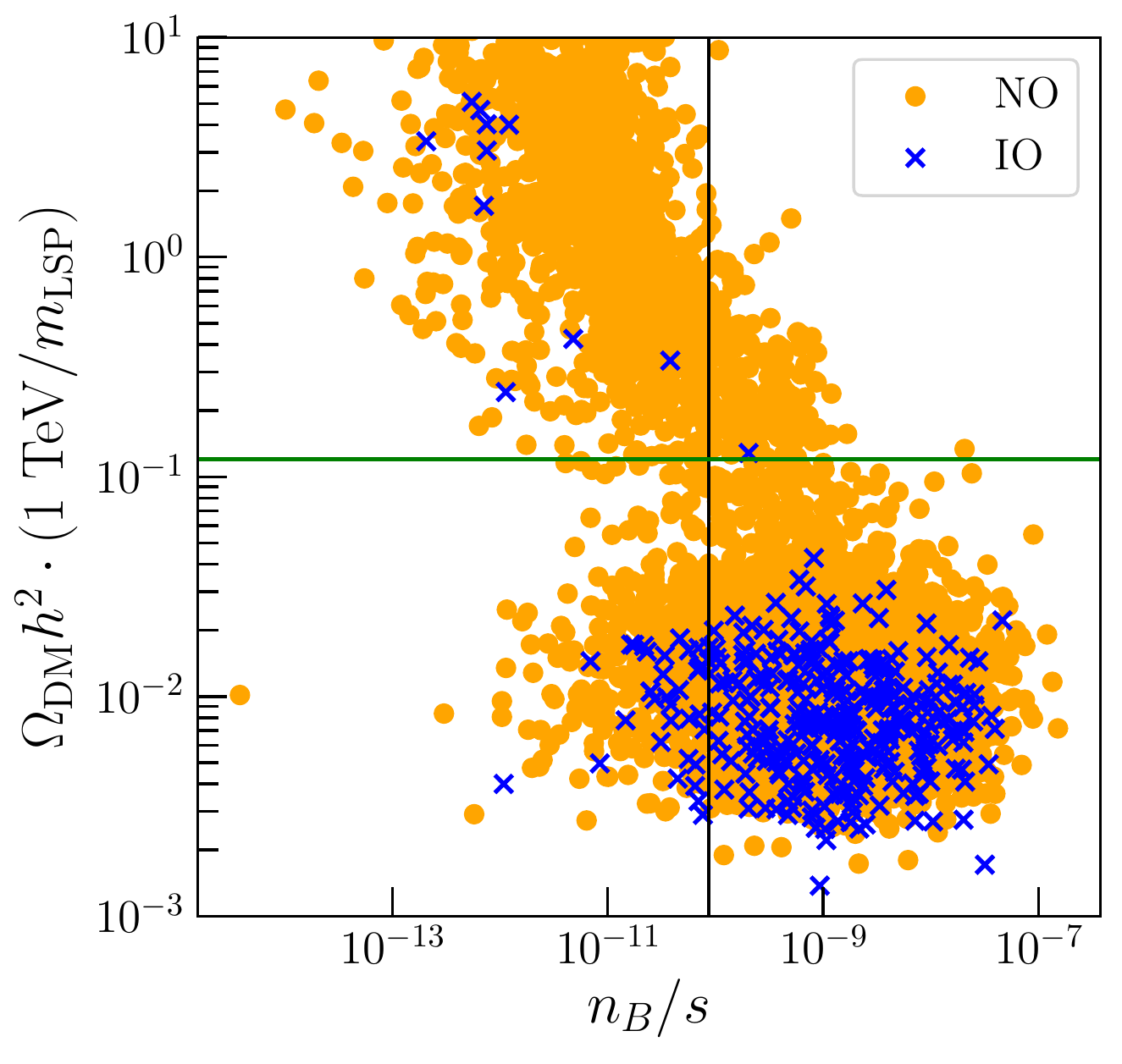}} 
\caption{
Non-thermal contribution to the LSP abundance vs $n_B/s$, with the
 observed values shown as the horizontal green and vertical black lines,
 respectively. 
}
\label{fig:bau_omdm}
\end{figure}

In Fig.~\ref{fig:bau_omdm}, we 
plot the non-thermal contribution to the LSP abundance from gravitino
decay
against the baryon asymmetry predicted at the same parameter point,
assuming $\Delta = 10^4$. The
vertical black and horizontal green lines show, respectively, the observed values of
baryon asymmetry and dark matter abundance $\Omega_{\rm DM} h^2 = 0.12$
\cite{Aghanim:2018eyx} for $m_{\rm LSP} = 1$~TeV.
We find that most of the points
predict $n_B/s \lesssim {\cal O}(10^{-9})$ and $\Omega_{\rm DM} h^2 \simeq
{\cal O}(10^{-2})$, where the typical values of $\lambda_6$ are
$\simeq {\cal O}(10^{-4})$ and $|\mu_1| \simeq |\mu_2| \simeq m_s$. 
The predicted value of $n_B/s$ is found to be larger than that estimated
in Refs.~\cite{egnno2, egnno3}; this is due to an enhancement in the
mass function $g(x)$ in Eq.~\eqref{eq:gx} for a degenerate mass
spectrum, which was neglected in the previous estimation. 
On the other
hand, we find many solutions
where the non-thermal component
of the LSP abundance from
gravitino decays accounts for the entire
dark matter density $\Omega_{\rm DM}
h^2 \simeq 0.12$. In this case, $\lambda_6 = {\cal O}(10^{-3})$, and the
singlet $\mu$ parameters are hierarchical, $m_s \ll |\mu_1| \ll
|\mu_2|$. For such parameter points, one must  ensure that the
thermal relic of the LSP is sufficiently depleted, which is obtained easily
if $\Delta \sim 10^4$, as we have assumed. 

There are also many solutions where the
abundance is found to be smaller than the observed value (particularly for IO). Therefore we expect the observed
dark matter abundance in these cases should be explained mainly by thermal relic LSPs. Notice, however, that the freeze-out density of the LSP can
be much larger than in a standard cosmological scenario due to the presence of the
dilution factor $\Delta$. This may revive a wide range of parameter
space in supersymmetric models where the thermal relic of the LSP would
otherwise be overabundant. A detailed study of this possibility will be given
elsewhere \cite{egnno5}.

In summary: we have examined the correlations between inflationary reheating, the non-thermal dark matter abundance produced by gravitino decays, neutrino masses, and the baryon asymmetry in a simple model based on a single master superpotential coupling $\lambda_6$ involving 
a gauge singlet, a heavy Higgs breaking
the GUT gauge symmetry and the (flipped) {\bf 10} matter representation. Using the known
neutrino mass-squared differences as a constraint, we find that the typical reheating temperature is 
$10^{12}$ GeV and the typical baryon-to-entropy ratio lies between $n_B/s \in (10^{-13} - 10^{-7})$, embracing the observed value near $10^{-10}$.
For the preferred value of the baryon asymmetry, we find that, for NO neutrino masses, the non-thermal LSP abundance may saturate the measured relic density of dark matter, but may be significantly lower, leaving open the possibility of a dominant thermal contribution. With IO masses, the
non-thermal component is typically subdominant. In this case, 
because of late entropy production, regions of parameter
space that would yield $\Omega_{\rm DM} h^2 \sim 1000$ in standard cosmology are preferred, opening new regions of supersymmetric parameter space for experimental searches.

\begin{acknowledgments}
\section*{Acknowledgments}
The work of J.E. was supported partly by the United Kingdom STFC Grant ST/P000258/1 
and partly by the Estonian Research Council via a Mobilitas Pluss grant. The work of M.A.G.G. was supported by the DOE grant DE-SC0018216. The work of N.N. was supported by the Grant-in-Aid for Young Scientists B (No.17K14270) and Innovative Areas (No.18H05542). The work of D.V.N. was supported partly by the DOE grant DE-FG02-13ER42020 
and partly by the Alexander S. Onassis Public Benefit Foundation. The work of K.A.O. was supported partly
by the DOE grant DE-SC0011842 at the University of Minnesota.

\end{acknowledgments}

\end{document}